\documentclass{PoS}
\usepackage{amsmath}
\usepackage{amssymb}
\usepackage{epsfig,graphicx}

\title{Open Issues in Neutrino Reactions}

\ShortTitle{Open Issues in Neutrino Reactions}

\author{\speaker{E. A. Paschos}\\%
       Department of Physics, TU-Dortmund, 44221 Germany\\
       E-mail: \email{paschos.e@gmail.com}}

\abstract{Numerous investigations discuss neutrino properties from low to ultra high energies. This review discusses several topics that are investigated in present day experiments. The first section covers the detection of $\tau$-neutrinos of cosmic origin and describes the significance of the results. The second topic points out that the reported reduction of the antineutrino flux in reactor experiments may be created by short wavelength oscillations where the detector averages over several wavelengths and observes only an average reduction of the antineutrino flux. In many theories the propagating particles are Majoranas and have an anapole moment; they scatter on atomic electrons with a distinct signature  in $\bar\nu_e+e^-\rightarrow\bar\nu_e+e^-$ scattering.}

\FullConference{Corfu Summer Institute 2016 "School and Workshops on Elementary Particle Physics and Gravity"\\
                 31 August - 23 September, 2016\\
                 Corfu, Greece}

\begin{document}

\section{Searching for Cosmic $\nu_\tau$'s}

The discovery of ultra-high energy cosmic neutrinos~\cite{Aartsen:2013jdh,Aartsen:2014gkd} opens a unique window to test neutrino properties and mixing at high energies~\cite{Beacom:2002vi}. Determining the lepton flavor composition of the flux is significant for understanding mechanisms of their generation, propagation and interaction. At the production point the ratios of flavors are the following,
\begin{equation}
\text{in pion and kaon decays:}\,\,\,\,\,\nu_e:\nu_{\mu}:\nu_{\tau}=1:2:0\label{eq:ratio_def}
\end{equation}
\begin{equation}
\text{in neutron decays:}\,\,\,\,\,\nu_e:\nu_{\mu}:\nu_{\tau}=1:0:0.\label{eq:ratio_def1}
\end{equation}

These ratios are modified in the propagation of states over cosmological distances. The oscillation between the states modifies the ratio among the three flavors generating a substantial component of $\tau$-neutrinos arriving on earth.

The large neutrino telescopes are tuned to detect $\tau$-neutrinos with results reported already by the IceCube collaboration. One analysis~\cite{Aartsen:2015ivb} computed the cross sections for the various flavors assuming similar energy and angular distributions for the three flavors and a functional form for the flux. The sum of contributions from the three flavors is compared with the observed number of events. Their results are shown in figure 1 with the colors indicating the degree of exclusion. Blue indicates the most excluded regions and red the preferred flavor composition on earth. The areas outside the curves are excluded at 68\% and 90\% confidence level. The black cross indicates the best fit for the composition on earth. This topic is just beginning to evolve and it is useful to consider additional criteria for identifying the component of tau neutrinos.

An alternative method for identifying $\tau$-neutrinos in a large range of energies uses the decay $\tau\rightarrow \mu\nu_\tau\bar\nu_\mu$. This can be done by identifying events which contain hadronic showers and long tracks from muons. The $\tau$-neutrino component is identified by measuring the energies $E_{shower}$ and $E_{\mu}$ and then plotting the events as a function of the ratio $y'=E_{shower}/(E_{shower}+E_{\mu})$. Since the leptonic decay of $\tau$  includes to missing neutrinos, the muon energy in the charged current interactions of $\nu_\tau$ is expected to be, on the average, lower. Alikhanov and myself carried out the detailed calculation~\cite{Alikhanov:2016qcj} by adopting the quark-parton model~\cite{Bjorken:1969in} since the reactions we consider take place at very high energies. For experimental reasons the events will be averaged over the cosmic flux from a minimum and a maximum energy and normalized to the total number of $\tau$-events in this range. We found

\begin{eqnarray}
\frac{1}{N^{(\nu+\bar\nu)}}\frac{dN^{(\nu+\bar\nu)}}{dy'}=\frac{1}{12(1+\text{Br}) y'^5}\nonumber\\ \times\left[\text{Br}\left(3y'^5-3y'^4+14y'^3-132y'^2+96y'\right)\right.\nonumber\\\left.-3 \text{Br} \left(y'^3-24y'^2+60y'-32\right)\log
   \left(1-y'\right)+9 y'^5 \left(y'^2-2 y'+2\right)\right],\label{eq:events34f}
\end{eqnarray}
where $\text{Br}=0.17$ is the branching ratio for $\tau\rightarrow \mu\nu_\tau\bar\nu_\mu$. We also took an average for incident neutrinos and antineutrinos, because in that case the quark distribution functions drop out in the ratios. In figure~2 is plotted the yield as a function of $y'$ for three incident fluxes $\nu_{\mu}:\nu_{\tau}=(1:0)$, $\nu_{\mu}:\nu_{\tau}=(1:1)$ and $\nu_{\mu}:\nu_{\tau}=(0:1)$. We note that for $y'>0.60$ the three cases can be distinguished.

The distribution of events demonstrates that more $\nu_\tau$-induced events populate the $y'>0.60$ region. In the same region the muon-induced events decrease monotonically as $y'\rightarrow1$. Thus it will be useful to determine the different shapes of the curves and separate the $\nu_\tau$ and $\nu_\mu$ components. Up to now the reported number of events do not suffice to carry out a statistical analysis, however more events will be accumulating in the coming years in the running and in new experiments.

\begin{figure}
\centering
\includegraphics[width=4.5in]{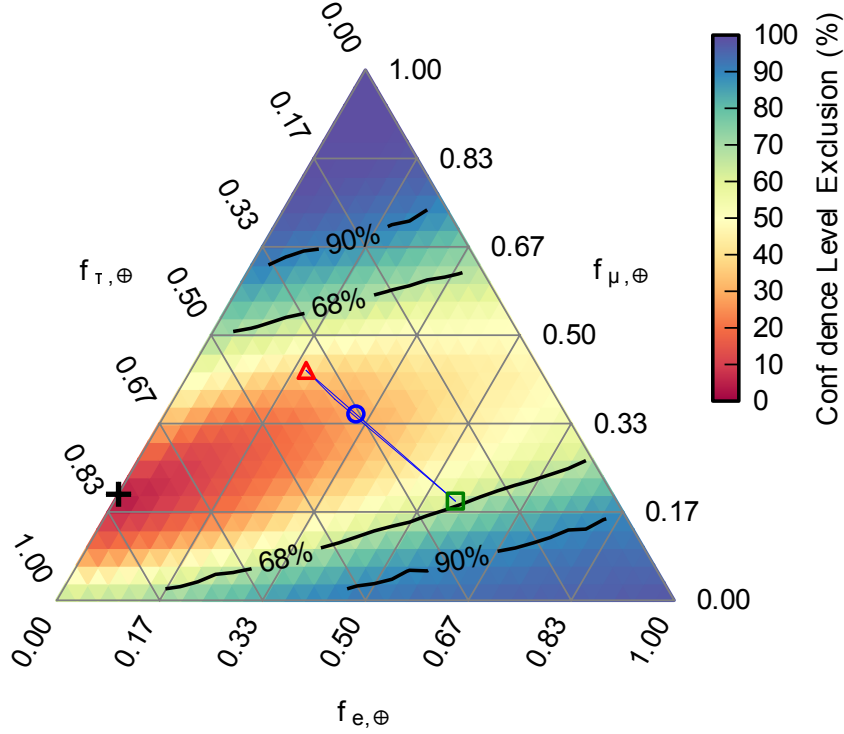}
\caption{Exclusion regions for astrophysical ratios ($f_e:f_\mu:f_\tau$) observed in the South Pole (from reference~\cite{Aartsen:2015ivb}).}
\label{s_chan1}
\end{figure}

\begin{figure}
\centering
\includegraphics[width=5.5in]{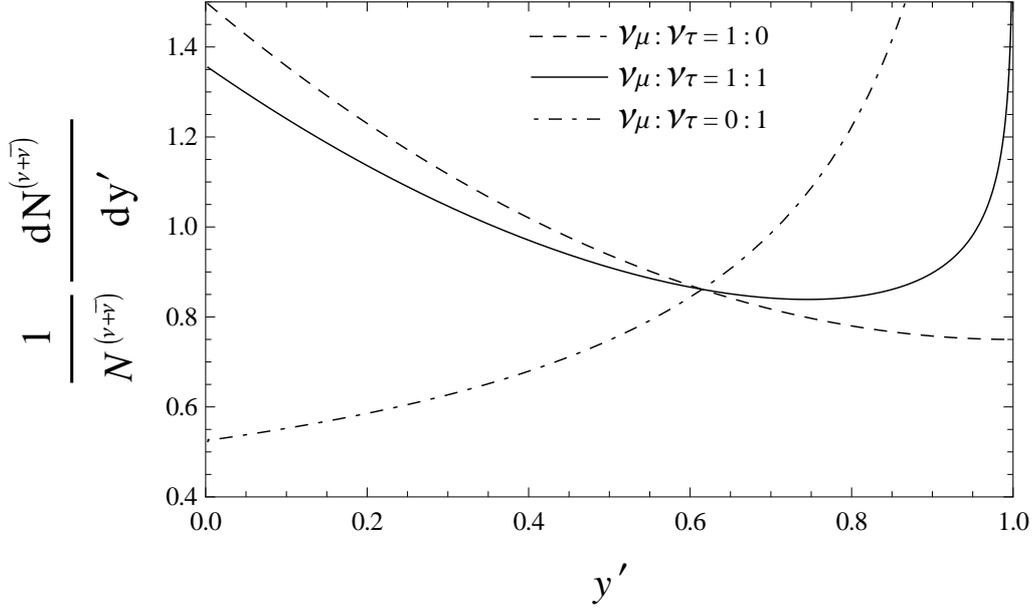}
\caption{Production spectra as a function of the visible inelasticity ($y'=E_{shower}/(E_{shower}+E_\mu)$). In the figure we show three flavor compositions of the neutrino flux: $\nu_{\mu}:\nu_{\tau}=1:0$ (dashed), $\nu_{\mu}:\nu_{\tau}=1:1$ (solid) and $\nu_{\mu}:\nu_{\tau}=0:1$ (dash-dotted).}
\label{s_chan2}
\end{figure}

\section{The reactor Anomaly}

In addition to the ultra-high energy neutrinos there are also several anomalies at lower energies which are explained by introducing additional states. The standard paradigm introduces three right-handed neutrinos being singlets under $SU(2)_L$. The general Lagrangian is written as

\begin{equation}
\mathcal{L}=\mathcal{L}_w+\mathcal{L}_s+\mathcal{L}_m, \label{lag_sum}
\end{equation}
where $\mathcal{L}_w$ is invariant under the electroweak group, $\mathcal{L}_s$ is invariant under the strong group and $\mathcal{L}_m$ is a mass term generated through a soft breaking of $SU(2)_L$. Denoting by $\nu_L=(\nu_e,\nu_\mu,\nu_\tau)$ and $N_R=(N_{R1},N_{R2},N_{R3})$ left- and right-handed neutrinos, respectively, we can write the mass term

\begin{equation}
\mathcal{L}_m=\frac{1}{2}\left(\begin{array}{cc}
\bar\nu_L & \bar N_R^C\\
\end{array}
\right)\left(\begin{array}{cc}
0 & m_D\\
m_D^T & M\\
\end{array}
\right)\left(\begin{array}{c}
\nu_L^C\\
N_R\\
\end{array}
\right)+h.~c. \label{eq:4}
\end{equation}

We seek solutions when the elements of $m_D$ are small relative to those of $M$ so that an expansion in $m_D/M$ is possible. The unitary matrix that diagonalizes the mass matrix is

\begin{eqnarray}
U=\left(\begin{array}{cc}
(1-\frac{1}{2}JJ^{\dag}) & J\\
-J^{\dag} & (1-\frac{1}{2}J^{\dag}J)\\
\end{array}
\right)\left(\begin{array}{cc}
U_{\theta} & 0\\
0 & U_{\chi}\\
\end{array}
\right)\label{eq:5}
\end{eqnarray}
with $J=m_D\dfrac{1}{M}$ and $U_{\theta}$ and $U_{\chi}$ diagonalize the submatrices $m_D\dfrac{1}{M}m_D^{T}$ and $M$, respectively. The Majorana mass terms split the Dirac neutrinos into pairs of Majorana particles. The light particles have masses of order $m_D\dfrac{1}{M}m_D^{T}$ and the heavy ones of order $M$.

For simplicity I restrict the discussion to two families of electrons and muons. The flavor states are given in terms of mass eigenstates $\psi_i(t)$

\begin{equation}
|\nu_e(t)\rangle=\sum\limits_{i=1}^{4}U_{ei}\psi_i(t)
\label{eq:7}
\end{equation}
with the coefficients given by

\begin{subequations}
\begin{align}
&U_{e1}=[1-\frac{1}{2}(JJ^{\dag})_{11}]c_{\theta}+\frac{1}{2}(JJ^{\dag})_{12}s_{\theta}\label{eq:10a}\\
&U_{e2}=[1-\frac{1}{2}(JJ^{\dag})_{11}]s_{\theta}-\frac{1}{2}(JJ^{\dag})_{12}c_{\theta} \label{eq:10b}\\
&U_{e3}=J_{11}c_{\chi}-J_{12}s_{\chi}\label{eq:10c}\\
&U_{e4}=J_{11}s_{\chi}+J_{12}c_{\chi},\label{eq:10d}
\end{align}
\end{subequations}
and similar formulas for $U_{\mu i}$ with $i=1,\ldots,4$~\cite{Paschos2016}.

I discuss next the reactor experiments where a beam of electron-type antineutrinos is generated and is detected at a distance of less than a few kilometers~\cite{Mueller:2011nm,Mention:2011rk}. In this situation there are three types of interference terms.

\noindent(i) The mixing of the two lighter eigenstates $\psi_1$ and $\psi_2$ is proportional to $\sin^2\left(\dfrac{\Delta m^2_{12}L}{4E}\right)$ which for distances of kilometers or less is very small and will be omitted.

\noindent(ii) The mixing between the two heavier states, $\psi_3$ and $\psi_4$, is of $O(J^4)$ and will also be neglected.

\noindent(iii) Finally there is the mixing of the heavier with the lighter states which produces the terms

\begin{equation}
P_{\bar\nu_e\rightarrow\bar\nu_e}=1-4\left\{|U_{e3}|^2+|U_{e4}|^2\right\}\sin^2\left(\frac{E_3-E_1}{2}L\right),\label{eq:15}
\end{equation}
where I assumed $E_3-E_1\approx E_4-E_1$. The energy differences are very likely large enough to produce oscillations with short wavelengths which cannot be resolved in the detectors. Thus they observe an average over several wavelengths with the consequence of replacing the sinus square term by $1/2$.

\begin{equation}
P_{\bar\nu_e\rightarrow\bar\nu_e}=1-2\left(|J_{11}|^2+|J_{12}|^2\right).\label{eq:16}
\end{equation}
This result is a candidate for the reactor and Gallium discrepancy. Attributing the decrease of the reactor fluxes to $-2\left(|J_{11}|^2+|J_{12}|^2\right)$ implies that either $|J_{11}|$ or $|J_{12}|$ or both are approximately $\sim0.10$.

In the formalism described in this section, the physical states $\psi_1,\ldots,\psi_4$ are Majorana neutrinos. Ordinary $\overset{\text{{\tiny (}}-\text{{\tiny )}}}{\nu_e}$  and $\overset{\text{{\tiny (}}-\text{{\tiny )}}}{\nu_\mu}$ beams produced in the laboratories oscillate according to equation~\eqref{eq:7} and develop Majorana components. The Majorana neutrinos have only two electromagnetic formfactors, an electric dipole moment and an anapole moment. We denote the electromagnetic current by $j_\mu(p_2,p_1)$ which interacts with the electromagnetic field $A_\mu$ as follows $H_{int}=A^\mu j_\mu$. We are looking to find consequences of this interaction without making any particular assumptions concerning the origin of the formfactors.

Let $u(p_1)$ and $u(p_2)$ be spinors of the initial and final neutrinos. The current is given in terms of two formfactors~\cite{Nowakowski:2004cv}

\begin{equation}
j^\mu(p_2,p_1)=\bar u(p_2)\left[i\varepsilon^{\mu\nu\alpha\beta}\sigma_{\alpha\beta}q_{\nu}F_3(q^2)+\left(q^\mu-\frac{q^2}{2m_\nu}\gamma^\mu\right)\gamma_5F_4(q^2)\right]u(p_1).\label{formf}
\end{equation}
$F_3(q^2)$ is the electric dipole moment which is odd under time-reversal and parity transformations. Up to now nobody measured $F_3(q^2)$ for any value of $q^2$. This is not surprising as we expect this formfactor  to be small by time reversal invariance. The second formfactor $F_4(q^2)$ was introduced by Zeldovich~\cite{Zeldovich} and is known as the anapole. In the non-relativistic limit the anapole interaction reduces to~\cite{Nowakowski:2004cv}

\begin{equation}
H^{NR}_{int}\propto F_{4}(0)\vec\sigma\cdot\left[\vec\nabla\times\vec B-\frac{\partial \vec E}{\partial t}\right].\label{anapole}
\end{equation}
Recalling the Maxwell equations, the term in the square brackets is proportional to electric current, i.e. in the absence of an electric current it vanishes. Consequently the anapole term contributes when it interacts with virtual photons. Under time reversal the anapole term is even but parity violating.

An attractive reaction to search for an anapole term is

\begin{equation}
\bar\nu_e+e^-\rightarrow\bar\nu_e+e^-.\label{reaction}
\end{equation}
In the scattering of the antineutrinos the recoiling electrons emerge close to the direction of the incoming particle (very small scattering angle).
The cross section differential in the electron energy has the form

\begin{equation}
\frac{d\sigma}{d E_{e}'}=\frac{G^2m}{4\pi}\left(\frac{F_4}{m_\nu}\right)^2\left[1+\left(1-\frac{E_e'}{E_\nu}\right)^2\right]\label{cross_s}
\end{equation} 
with $m$ the mass of the electron and $E'_e$ the energy of the recoiling electron. The cross section decreases monotonically as $E_e'\rightarrow E_\nu$ but not as fast as background. The background originates from charged and neutral current reactions of the electroweak theory. This cross section is

\begin{equation}
\frac{d\sigma}{d E_{e}'}=\frac{G^2m}{4\pi}\left[\frac{1}{4}+2.25\left(1-\frac{E_e'}{E_\nu}\right)^2\right],\label{cross_sw}
\end{equation} 
where I replaced the weak couplings $g_V$ and $g_A$ by their values. It also decreases as $E_e'\rightarrow E_\nu$ by a larger factor and reaches the limiting value $\dfrac{G^2m}{16\pi}$. The TEXANO experiment~\cite{Deniz:2009mu,Deniz:2010mp} observed the decrease. However, the experimental errors are still very large to determine the limiting value or possible deviations (see figure 16 in reference~\cite{Deniz:2009mu}). The cross section in equation~\eqref{cross_s} appears smaller but we should keep in mind that the factor $(F_4/m_\nu)^2$ is still unknown and can assume a large value. A similar analysis is possible for the reaction $\nu_\mu+e^-\rightarrow\nu_\mu+e^-$ and also in neutrino scattering on the Coulomb field of nuclei. A careful analysis must include also precise estimates of the neutrino flux.

\section{Summary}

Neutrino telescopes with large volumes of water began detecting astrophysical neutrinos. The flavor content of the cosmic flux is of great interest and I described a new method to identify the $\nu_\tau$ component.

In the low energy experiments, models with right-handed neutrinos always develop into Majorana particles whose interactions may become evident in electromagnetic interactions. Their scattering on atomic electrons will be a direct signal for identifying them. For such states I describe properties and results from the anapole formfactor.

An anapole moment has been considered as a coupling to WIMPs~\cite{Pospelov:2000bq} and other dark matter particles~\cite{Ho:2012bg}. I considered here the possibility that Majorana neutrinos when they propagate, they interact through the anapole moment. The new coupling will modify the recoil spectra of electrons for the reaction in equation~\eqref{reaction}. An analysis of neutrino interactions with general couplings appeared after the conference~\cite{Rodejohann:2017vup}. Their comparisons agree with the standard model calculations. Thus the studies so far indicate that we need accurate  data for neutrino--electron elastic scattering. New results will be coming from present experiments~\cite{Park:2015eqa}, from beam-dump experiments~\cite{Alekhin:2015byh} and from the European Spallation Source~\cite{Baussan:2013zcy}.

\acknowledgments
I wish to thank Dr. I. Alikhanov for discussions and help in preparing the manuscript.

\end{document}